# 30°-twisted bilayer graphene quasicrystals from chemical vapor deposition

Sergio Pezzini,[1, 2, *] Vaidotas Mišeikis,[1, 2] Giulia Piccinini,[1, 3] Stiven Forti,[1] Simona Pace,[1, 2] Rebecca Engelke,[4] Francesco Rossella,[3, 5] Kenji Watanabe,[6] Takashi Taniguchi,[6] Philip Kim,[4] and Camilla Coletti[1, 2, *]

[1]*Center for Nanotechnology Innovation @NEST, Istituto Italiano di Tecnologia, Piazza San Silvestro 12, 56127 Pisa, Italy*

[2]*Graphene Labs, Istituto Italiano di Tecnologia, Via Morego 30, 16163 Genova, Italy*

[3]*NEST, Scuola Normale Superiore, Piazza San Silvestro 12, 56127 Pisa, Italy*

[4]*Department of Physics, Harvard University, Cambridge, MA 02138, USA*

[5]*NEST, Istituto Nanoscienze-CNR, Piazza San Silvestro 12, 56127 Pisa, Italy*

[6]*National Institute for Materials Science, 1-1 Namiki, Tsukuba, 305-0044, Japan*

ABSTRACT. The artificial stacking of atomically thin crystals suffers from intrinsic limitations in terms of control and reproducibility of the relative orientation of exfoliated flakes. This drawback is particularly severe when the properties of the system critically depends on the twist angle, as in the case of the dodecagonal quasicrystal formed by two graphene layers rotated by 30°. Here we show that large-area 30°-rotated bilayer graphene can be grown deterministically by chemical vapor deposition on Cu, eliminating the need of artificial assembly. The quasicrystals are easily transferred to arbitrary substrates and integrated in high-quality hBN-encapsulated heterostructures, which we process into dual-gated devices exhibiting carrier mobility up to $10^5$

cm$^2$/Vs. From low-temperature magnetotransport, we find that the graphene quasicrystals effectively behave as uncoupled graphene layers, showing 8-fold degenerate quantum Hall states: this result indicates that the Dirac cones replica detected by previous photo-emission experiments do not contribute to the electrical transport.

Twisted bilayer graphene (TBG), a system made of two stacked single-layer graphene (SLG) with misaligned crystallographic orientation, is providing an incredibly rich platform for novel physical phenomena [1,2]. In the limit of small twist angle (<10°), the long-range moiré superpotential determines important modifications to the structural and electronic properties [3-5]. For larger twisting, TBG falls in a weak coupling regime, with the electronic properties resembling those of two SLG conducting in parallel [6, 7]. However, considerable interlayer coupling seems to re-emerge in TBG with a twist angle of precisely 30 degrees, giving rise to multiple Dirac cones replica, as detected by angle-resolved photoemission spectroscopy (ARPES) in two recent experiments [8, 9]. In this configuration, TBG forms an incommensurate structure with 12-fold rotational symmetry – although lacking of translational symmetry – known as dodecagonal quasicrystal (QC). In addition to the cones' multiplication, QC-TBG is predicted to host spiral Fermi surfaces resulting in novel quantum oscillations [10], van Hove singularities [11,12] and semi-localized electronic states following the dodecagonal tiling [11], which strongly motivates further experimental investigation, in particular low-temperature electrical transport in high-mobility devices. However, QC-TBG is extremely sensitive to small variation in the twist angle, which makes it highly challenging to realize it with the common hBN-mediated "tear-and-stack" approach [13]. The samples presented by Anh *et al.* [8] and Yao *et al.* [9] were grown on SiC and Pt(111) substrates, respectively, which pose severe limitation to the possibility of transfer and processing high-mobility transport devices. Chemical vapor deposition (CVD) of single-crystalline graphene bilayer on Cu foils offers an appealing alternative approach, as it allows to obtain large-area defect-free samples that can be isolated from the growth catalyst and encapsulated in hBN flakes with several methods [14, 15]. Moreover, a few works have suggested the possibility of a highly selective growth mechanism, strongly favoring the AB and 30°-rotated stacking over random twisting on Cu(111) foils [16] and Cu/Ni gradient alloy [17]. We found that our CVD process for isolated graphene single crystals [18, 19] yields large-area bilayers with AB or QC-TBG configurations exclusively. The two types can be easily distinguished on the basis of the relative orientations of the hexagonal edges of the two layers, allowing for large-scale production of QC-TBG, transfer to arbitrary substrates and device processing. The QC-TBG configuration is unambiguously identified by transmission electron microscopy (TEM) measurements on suspended samples and supported by Raman spectroscopy. We fabricated hBN-encapsulated dual-gated devices and probed for the first time the low-temperature magnetotransport properties of QC-TBG.

To produce the graphene crystals, we made use of electropolished high-purity Cu foils (~2 $cm^2$) and performed low-pressure CVD in a commercial cold-wall reactor, limiting the gas flow via a quartz enclosure (more details can be found in SI and Refs.[18, 19]). We obtained isolated hexagonal graphene domains with 150-250 μm lateral dimensions, with additional hexagonal concentric bilayer regions with 10-50 μm lateral dimensions, sharing the same nucleation center as the monolayer portion (see Figure 1a, b). Dark field optical imaging on the Cu foil allows easy identification of the graphene flakes and discerning of the stacking order of the bilayer region.

Over a representative growth cycle, we observed a total of 114 graphene single crystals with bilayer portions. Among them, 59% showed AB stacking (confirmed by Raman spectroscopy, see Figure S1) and 41% showed 30° rotation; no other stacking was identified by optical inspection. While AB stacking is energetically favored by interlayer van der Waals interaction [20] – and therefore dominates in natural graphite crystals – the incommensurate QC-TBG (Figure 1c) can be stabilized by the surface morphology of the growth substrate. Yan *et al.* [16] showed that the interaction between step edges on Cu(111) and the graphene layers at the nucleation stage is optimal for straight (zigzag or armchair) graphene edges oriented along the step direction. This preference applies to repeated sequential nucleation in multilayers growth and makes the 0° and 30° twisting (approximately) isoenergetic, favoring them over random angles. In our case, graphene is synthetized over large predominantly (111) domains, obtained with short-time annealing (more details in SI and Ref.[18]), where AB stacked and QC bilayers are equally probable according to Ref.[16]. In this sense, additional control on the stepped surface morphology via fine-tuning of the annealing parameters is not expected to increase significantly the occurrence of QC-TBG with respect to that of AB bilayers. Although additional multilayer patches can attain non-negligible size for longer growth time, they can be easily identified and avoided during further characterization and processing of the QC-TBG. To isolate the CVD-grown QC-TBG from the Cu catalyst, we employed a semi-dry PMMA-mediated transfer technique, based on electro-chemical delamination from Cu [21] and deterministic lamination onto an arbitrary substrate (more details can be found in SI and Ref.[19]). In Figure 1d we show a typical graphene crystal with 30°-rotated bilayer portion after transfer to $SiO_2$ (285 nm)/Si, showing no visible wrinkles, cracks or macroscopic contamination. We then adopted the procedure described in Refs.[15, 22] to pick up the QC-TBG from $SiO_2$ with hBN flakes (10-50 nm thick) and assemble encapsulated structures with large contamination-free areas (see Figure S2 for optical and atomic force microscopy images). Straight edges of the hBN flakes were intentionally misaligned with respect to the two hexagonal layers, ideally falling at 15° misorientation, minimizing the chance of an hBN-induced moiré superstructure. Moreover, despite the fact that high temperatures favor the interface cleaning process [22], the assembly was performed keeping $T < 170$ °C to avoid possible motion of the encapsulated graphene sheets [23] and relaxation of the TBG towards AB stack [24].

In Figure 1e we show a transmission electron microscope (TEM) selected area electron diffraction (SAED) pattern measured on a QC-TBG suspended over a 2.5 µm diameter hole in a $Si_3N_4$ membrane, obtained via the same transfer method described above. The diffraction pattern consists of two sets of the monolayer graphene Bragg peaks, rotated at 30° relative to one another, resulting from the two (top and bottom) layers. The rotation angle is measured to be within 0.1° of 30°, given the uncertainty of the measurement (see Figure S3 for details). Likely more robust against slight image distortions, we observe 12-fold rotational symmetry in the fact that the $11\bar{2}0$ Bragg peaks from each layer are radially aligned with the $10\bar{1}0$ and $20\bar{2}0$ peaks from the other layer. This result was confirmed over numerous suspended QC-TBG regions (see Figure S3), proving high uniformity in the twist angle over distances of few tens of microns. In addition, in Figure S4, we present SAED on hBN-encapsulated QC-TBG, which confirms that the system maintains the growth-determined 12-fold rotational symmetry and interlayer rotation (within 0.2° of 30°) when subjected to van der Waals assembly. These observations identify our CVD graphene bilayers as dodecagonal quasicrystals and open the possibility of more in-depth studies of the electronic response of this system.

With this important piece of information in hand, we moved to a characterization of the QC-TBG with Raman spectroscopy (see SI for technical details). In Figure 2a we compare representative Raman spectra of CVD grown SLG, QC-TBG and hBN-encapsulated QC-TBG, all of them placed on $SiO_2$/Si. While AB bilayer graphene has a clear Raman fingerprint in the form of a multi-Lorentzian 2D peak [25] (see Figure S1), large-angle TBG provides the same Raman response as SLG, i.e. sharp G and 2D peaks, although with minor rotation-dependent evolution and features [26, 27, 28, 29]. In our QC-TBG samples, we could resolve a double peak structure in the low-energy 1300-1400 $cm^{-1}$ region, which is completely absent for the surrounding SLG portions, as shown in Figure 2b. While for SLG this range is occupied by the defect-induced D band [25] (although absent in our SLG, indicating a negligible defect concentration), large-angle TBG is known to exhibit a D-like band [29] accompanied by an intervalley double-resonance R mode induced by the potential mutually imposed by the two layers [26]. In Figure 2c we present a scanning Raman map of the Intensity of the R mode over the TBG region of the sample shown in Figure 1d, which serves as a high-contrast alternative imaging approach for the QC-TBG. The R peak here is positioned at 1374($\pm$6) $cm^{-1}$, in good accordance with the recent findings by Yao *et al.* [9], who identified it as due to scattering with a reciprocal lattice vector of the second layer, a process specific to QC-TBG. As for the hBN-encapsulated QC-TBG, this peak could not be identified, due to its overlap with the high-intensity $E_{2g}$ mode from the thick hBN crystals.

In Figure 2d-f we present comparative correlation plots of the full width at half maximum of the 2D peak FWHM(2D), intensity ratio of the 2D and G peaks I(2D)/I(G), and frequency of the 2D peak $\omega_{2D}$, respectively, as a function of the frequency of the G peak $\omega_G$. The evolution of these characteristics from SLG to QC-TBG (gray to blue points) qualitatively matches the early report of Kim *et al.* for large-angle TBG [27]: FWHM(2D) drastically decreases, reaching values typically observed in hBN-encapsulated SLG [30]; I(2D)/I(G) doubles, accompanied by a redshift of $\omega_G$, indicating a reduction in the doping level [31], which we ascribe to charge redistribution among the two graphene layers; $\omega_{2D}$ blueshifts by 10 cm$^{-1}$, possibly due to a reduction of the global strain [32], modification of the dielectric environment [33] or reduction of the Fermi velocity [34]. Upon dry encapsulation of QC-TBG in hBN (blue to red points), we found that FWHM(2D) is reduced to as low as 13 cm$^{-1}$, which is, to the best of our knowledge, the lowest value reported for hBN-encapsulated graphene, suggesting a drastic reduction of strain fluctuations and possibly a high-mobility electronic system. Moreover, the large increase of I(2D)/I(G) indicates low doping and therefore the possibility of probing this system in the vicinity of the Dirac crossings. The further blueshift of $\omega_{2D}$, finally, matches the findings of Ref.[33] regarding dielectric screening from hBN.

Starting from our hBN-encapsulated QC-TBG, we fabricated edge-contacted double-gated Hall bar devices for magnetotransport measurements (see sketch in Figure 3d inset and Figure S5). Figure 3 summarizes the main findings for high mobility device D1, while analogous results from the lower mobility device D2 are presented in Figure S6. In Figure 3a we show the resistivity of D1 as a function of the carrier density $n$ (black curve), measured at $T = 4.2$ K. By combining the top and bottom gate potentials $V_{tg}$ and $V_{bg}$, we can modulate $n$ while keeping the interlayer displacement field $D$ fixed (in this case to $D = 0$, meaning an equal charge distribution between the two layers). In a double-gate configuration, $n = 1/e \, (C_{tg}V_{tg} + C_{bg}V_{bg}) - n_0$ and $D = (C_{tg}V_{tg} - C_{bg}V_{bg})/2 - D_0$, where $C_{tg}$ and $C_{bg}$ are the capacitance per unit area of the top and bottom gate, while $n_0$ and $D_0$ are residual carrier density and displacement field at zero gate voltages [6]. From gate-dependent Hall effect measurements of D1, we estimate $C_{tg} = 1.76 \times 10^{-7}$ F/cm$^2$, $C_{bg} = 1.05 \times 10^{-8}$ F/cm$^2$ and $n_0 = 2.6 \times 10^{11}$ cm$^{-2}$, while $D_0/\varepsilon_0 = 0.047$ V/nm is obtained by inspecting the resistivity oscillations. The narrow resistivity peak ($n^* = 3 \times 10^{10}$ cm$^{-2}$) and carrier mobility of $10^5$ cm$^2$/Vs (red circles) are comparable to state-of-the-art devices based on CVD graphene [14, 15], assuring a low-disorder platform where the intrinsic electronic properties of the dodecagonal quasicrystal can be accessed. In Figure 3b we show a map of the first derivative of the Hall conductivity $d\sigma_{xy}/dV_{tg}$ as a function of the two gate potentials, taken at fixed $B = 1$ T applied perpendicular to the device plane. Already at this moderate field, we can fully resolve an intricate pattern, with zeroes in $d\sigma_{xy}/dV_{tg}$ (black areas) corresponding to quantum Hall states, separated by a number of crossings between Landau levels (LLs). In Figure 3c we plot the same data as a function of the filling factor $v = 1/B \times h/e \times n$ and $D/\varepsilon_0$, where $n$ and $D$ are obtained via the relations specified above. Aligned along $D = 0$ (red dashed line), we observe a series of nodes separating quantum Hall states positioned at $v = -4, -12, -20,\ldots$ , while at higher displacement field (blue dotted line) additional states arise at $v = -8, -16, -24,\ldots$ . According to Sanchez-Yamagishi *et al.* [6], this phenomenology is understood in terms of crossings between LLs belonging to independent graphene layers, controlled by modulation of $n$ and $D$ via the two gates. Across this pattern, the Hall conductivity $\sigma_{xy}$ is expected to vary in the form of quantized steps $N \times e^2/h$, with $N$ corresponding to the degeneracy of the LLs. In Figure 3d we show $\sigma_{xy}$ as a function of $n$ for $D = 0$ (red line), and $D \gg 0$ (blue line, taken along the blue dotted line in Figure 3c) demonstrating two series of equally spaced plateaus at $-4, -12, -20, \ldots \times e^2/h$ and $-8, -16, -24,\ldots \times e^2/h$, respectively. The degeneracy $N = 8$ is due to spin, valley and layer degree of freedom, as in the case of two in-parallel uncoupled SLG [6]. If Dirac cones replica contributed to the density of states in QC-TBG, one would expect $N$ in excess of 8. However, our data indicate the absence of additional degeneracies, ruling out this scenario. In addition, strong interlayer coupling should lead to anomalous screening properties of QC-TBG with respect to the independent SLG case [8]. We can directly probe the response of the system to field-effect doping by inspecting the positioning of the LLs crossing as a function of $D$ (see Figure S7). Each crossing corresponds to a known difference in charge density and electrostatic potential between the two graphene sheets, from which we estimate an interlayer capacitance per unit area $C_{gg}$ = 6.9 ± 1.4 µF/cm$^2$, in agreement with the reports of Ref.[6,35]. When measuring a reference device, D3, made by artificial stacking of 30.2°-misaligned mechanically exfoliated graphene flakes [13], we found completely analogous results (see Figure S8), thus confirming the uncoupled layers scenario. We investigated further the Fermi surface of QC-TBG, by measuring the resistance of D1 as a function of $B$ (up to 10 T) and $n$, both for null and finite $D$ (Figure 3e, left and right panel, respectively). For $D$ = 0 we observed a standard fan diagram, with the oscillatory component of the resistance at fixed $n$ showing a single frequency $f_0$ as a function of $B^{-1}$, as demonstrated by the fast Fourier transform spectra (FFT, see Figure 3f, left). We found that $f_0$ varies as a function of $|n|$ with slope $h/e \times 1/8$ (dashed blue line), which indicates a single 8-fold degenerate Fermi surface [36]. Applying a displacement field $D/\varepsilon_0$ = -0.14 V/nm (Figure 3e, right) is sufficient to lift the layers' degeneracy, resulting in a rather complex fan with numerous level crossings as a function of $B$ and $n$. The corresponding FFT spectral map (Figure 3f, right) shows multiple peaks, with the two main frequencies $f_1$ and $f_2$ resulting from two layer-resolved Fermi surfaces, and the sum frequency $f_1+f_2$ showing approximately double slope as compared to $f_0$, as expected for 4-fold degeneracy. On this basis, and with analogous data obtained on device D2, we can confirm that no Fermi contours, other than the ones surrounding the **K** and **K'** points of the two layers, contribute to the magnetotransport of QC-TBG.

Since the incommensurate dodecagonal quasicrystal emerges only at exact 30° rotation, it is important to discuss how variations in the twist angle within our experimental uncertainty (see Figure S3-S4) can affect the electrical transport properties. So-called approximants of QC-TBG, i.e. TBG with translational symmetry obtained by slight angular deviations from 30° [11] or changes in the lattice constant of one of the two layers [12], show a smooth evolution of the density of states as a function of angle [11] and can accurately describe the photoemission results [12]. Based on those calculations, one expects the transport properties to be robust within tenths-of-degree variations (comparable to our experimental uncertainty) and, therefore, to be representative of QC-TBG ones. Extrinsic scattering of the charge carriers might also hinder the observation of anomalous interlayer coupling via electrical transport. Sample D1, however, shows very high carrier mobility at low temperature, ensuring a mean free path in the order of the channel width (1 µm). While D2 surely incorporates more disorder, the results on the two devices do not differ significantly (e.g. one just needs to apply different magnitudes of magnetic and electric fields to resolve equivalent LLs patterns). In comparison to our hBN-encapsulated samples, which show minimal residual doping, the ones used in Ref.[8] present considerable coupling to the growth substrate. This is evidenced by a Fermi level of 0.3 eV for both graphene layers and a layer-asymmetric charge transfer dynamic [37]. A possibility is that some of the ARPES results are specific to the SiC-supported system and/or to the highly doped regime. In this sense, we note that the samples of Ref.[9] are less doped and the authors report much weaker and limited-in-number Dirac cones replica. Nevertheless, in our experiments, we did not observe any anomaly as a function of field-effect doping up to 0.17 eV (corresponding to the largest carrier density measured in D2 at $D = 0$, calculated taking into account the layer degeneracy). Finally, one should consider the possibility that the extra Dirac cones are so-called final states, i.e. purely spectroscopic features due to scattering of electrons at the surface *after* the photo excitation. Such elements, known both for single-layer graphene on SiC [38] and aligned graphene/hBN [39], do not correspond to additional superpotential-induced states and, therefore, are not involved in the electrical transport: under this scenario, our results are consistent and complementary to those of Ref.[8,9].

In conclusion, we show that large-area graphene bilayers with 30° interlayer rotation can be grown by chemical vapor deposition on Cu. This approach has several advantages, in particular allowing for integration of the QC-TBG into hBN-encapsulated structures. Despite the PMMA-mediated transfer process, these samples exhibit extremely low doping and strain fluctuations, as shown by a Raman 2D peak as narrow as 13 cm$^{-1}$. The quantum Hall regime, fully developed at $B \leq 1$ T, presents gate-tunable 8-fold degenerate states, as for the standard case of uncoupled graphene layers. While QC-TBG probed by ARPES can show Dirac cones' replica with dodecagonal symmetry [8], we have demonstrated that the low-energy band structure and density of states - the physical properties probed by magnetotransport - are unaffected by this mechanism. During the review process, we became aware that Deng *et al.* [40] have come to similar conclusions via scanning tunneling spectroscopy of QC-TBG near the Dirac point. Theoretical calculations, however, suggest important modification due to interlayer coupling when populating the bands up to few eV [11,12], a regime not accessible with standard field-effect devices, calling for alternative experimental techniques. In more general terms, our work demonstrates an alternative approach to high quality electrical transport experiments in twisted graphene layers, where the twist angle is precisely determined during rapid CVD growth of a macroscopic number of large-area samples, without imposing limitations in terms of carrier mobility. While at present the twist angle cannot be arbitrarily selected such as in the tear-and-stack approach, advancements in the control of the Cu surface morphology might enable to obtain arbitrary and controlled deviations from both the AB and 30°-rotated stacking orders.

FIGURES.

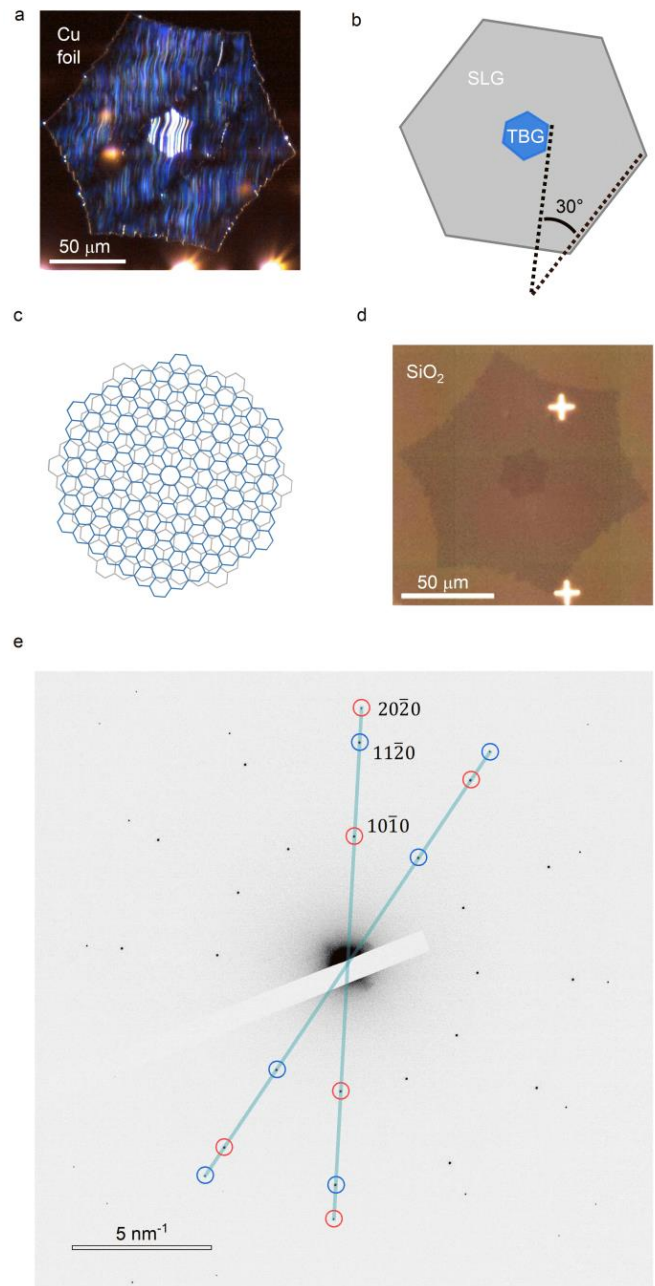

**Figure 1.** (a) Dark field optical microscopy image of a typical QC-TBG grown by CVD on Cu foil. (b) Schematic structure of the sample shown in (a), with the two superimposed concentric hexagons rotated by 30°. (c) Sketch of the crystal structure of the 12-fold rotationally symmetric QC-TBG. (d) Bright field microscopy image of a QC-TBG after transfer by semi-dry method on a SiO$_2$ (285 nm)/Si wafer. (e) TEM SAED obtained on a QC-TBG suspended over a TEM grid. The red (blue) circles identify the Bragg peaks from the top (bottom) graphene layers. Three groups of spots are detected and labelled according to their Miller indices. The straight lines are guides to the eye showing the radial alignment of peaks form different layers.

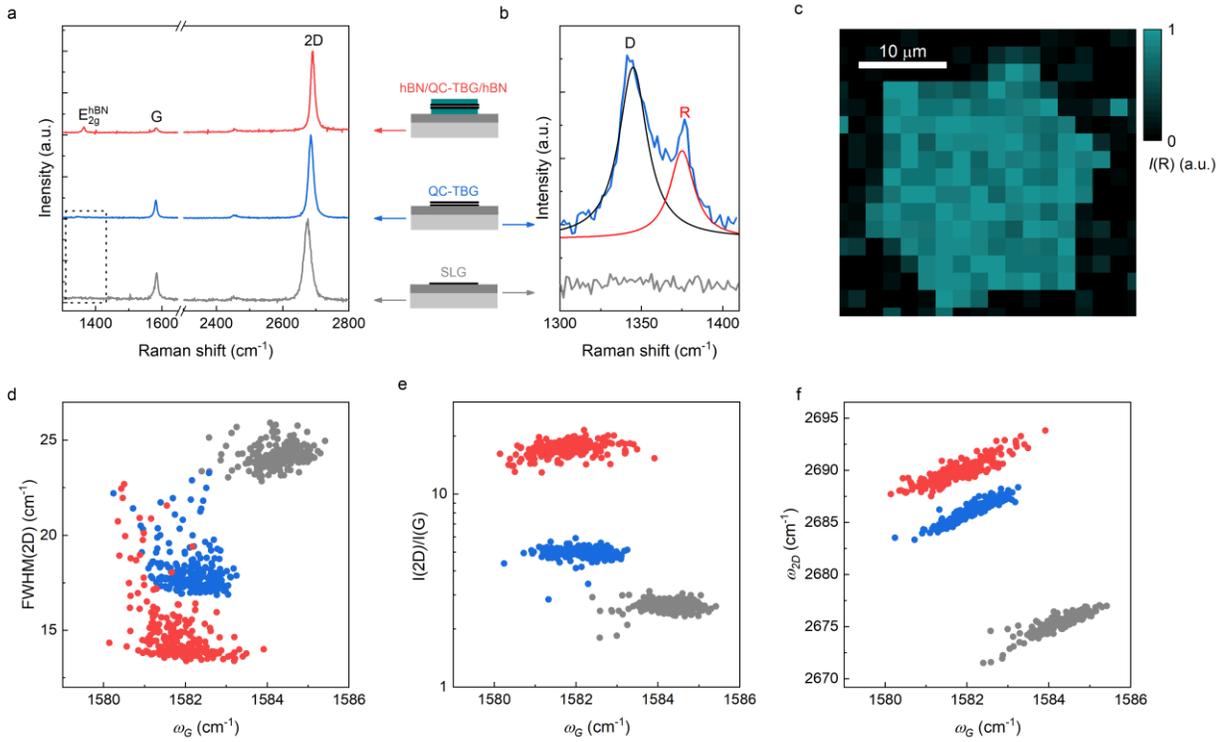

**Figure 2.** (a) Representative Raman spectra of SLG on SiO$_2$/Si (grey), QC-TBG on SiO$_2$/Si (blue), hBN-encapsulated QC-TBG (red). All the samples are CVD grown on Cu and then transferred (and assembled) as described in the text. The sketches on the right-hand side identify the three structures considered (light grey = Si; dark grey = SiO$_2$; black = graphene; dark cyan = hBN). (b) Enlarged view of the low-energy region for SLG and QC-TBG on SiO$_2$/Si; the acquisition time was increased by ×20 with respect to the spectra in (a). The black and red lines are Lorentzian fits to the D and R modes of QC-TBG. (c) Scanning Raman mapping of the intensity of the R mode over the TBG area of the sample in Figure 1(d). (d-f) Correlation plots of FWHM(2D), I(2D)/I(G)

and $\omega_{2D}$ as a function of $\omega_G$ from 15×15 μm² areas of SLG on SiO$_2$/Si (grey), QC-TBG on SiO$_2$/Si (blue), hBN-encapsulated QC-TBG (red). Each point corresponds to a single spectrum, acquired at 1 μm interval.

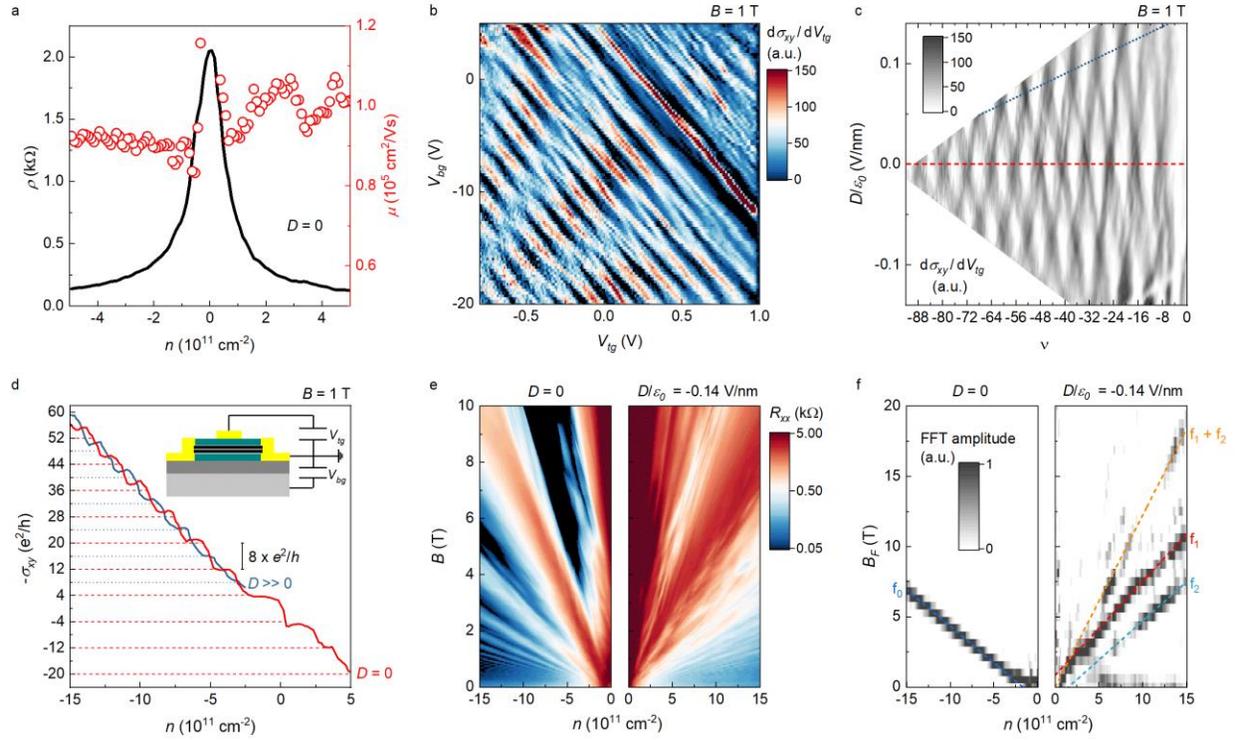

**Figure 3.** (a) Resistivity (black line) and carrier mobility (red circles) of sample D1 as a function of carrier density, at zero displacement field. (b) First derivative of $\sigma_{xy}$ as a function of $V_{tg}$ and $V_{bg}$, at $B = 1$ T. (c) Same data of (b) as a function of $v$ and $D/\varepsilon_0$. The red dashed line marks $D = 0$. The large-$D$ data in the following panel are taken along the blue dotted line. (d) Hall conductivity as a function of carrier density at $B = 1$ T, for $D = 0$ (red line) and $D \gg 0$ (blue line). The inset shows a schematic cross section of the devices studied (light grey = Si; dark grey = SiO$_2$; black = graphene; dark cyan = hBN; yellow = Cr/Au). (e) Longitudinal resistance as a function of carrier density for zero (left panel) and finite (right) displacement field. (f) Fast Fourier transform of the oscillatory component of the resistance data in (e). The dashed lines $f_{0,1,2}$ are fits the main peaks in the FFT spectra as a function of $n$, while $f_1+f_2$ is the sum of the fits to the two splitted components. All the data were acquired at $T = 4.2$ K.

ASSOCIATED CONTENT

**Supporting Information.** Details on the experimental methods for CVD growth, semi-dry transfer from Cu to $SiO_2$/Si, encapsulation in hBN, TEM, Raman spectroscopy, device fabrication and low-temperature magnetotransport. Figures S1-S8.

AUTHOR INFORMATION

**Corresponding Authors**

[*] sergio.pezzini@iit.it; camilla.coletti@iit.it

**Notes**

The authors declare no competing financial interest.

ACKNOWLEDGMENT

The authors thank Pilkyung Moon and Vladimir Fal'ko for useful discussions. Growth of hexagonal boron nitride crystals was supported by the Elemental Strategy Initiative conducted by the MEXT, Japan and the CREST(JPMJCR15F3), JST. The research leading to these results has received funding from the European Union's Horizon 2020 research and innovation program under grant agreement No. 785219-GrapheneCore2 and 881603-GrapheneCore3. The work at Harvard is supported by ONR (N00014-15-1-2761).

*Supporting Information for*

30°-twisted bilayer graphene quasicrystals from chemical vapor deposition

Sergio Pezzini,[1, 2, *] Vaidotas Mišeikis,[1, 2] Giulia Piccinini,[1, 3] Stiven Forti,[1] Simona Pace,[1, 2] Rebecca Engelke,[4] Francesco Rossella,[3, 5] Kenji Watanabe,[6] Takashi Taniguchi,[6] Philip Kim,[4] and Camilla Coletti[1, 2, *]

[1]*Center for Nanotechnology Innovation @NEST, Istituto Italiano di Tecnologia, Piazza San Silvestro 12, 56127 Pisa, Italy*

[2]*Graphene Labs, Istituto Italiano di Tecnologia, Via Morego 30, 16163 Genova, Italy*

[3]*NEST, Scuola Normale Superiore, Piazza San Silvestro 12, 56127 Pisa, Italy*

[4]*Department of Physics, Harvard University, Cambridge, MA 02138, USA*

[5]*NEST, Istituto Nanoscienze-CNR, Piazza San Silvestro 12, 56127 Pisa, Italy*

[6]*National Institute for Materials Science, 1-1 Namiki, Tsukuba, 305-0044, Japan*

**CVD growth**

QC-TBG samples are grown on electropolished Cu foil by chemical vapour deposition in a cold-wall CVD reactor (Aixtron 4" BM-Pro) at a pressure of 25 mbar and a temperature of 1060 °C. The sample is placed inside a quartz enclosure to limit the gas flow on the sample and control the nucleation density. Initially, the foil is annealed in pure Ar atmosphere at the growth temperature for 10 minutes, followed by 20 minutes of growth in an atmosphere comprised of 90% Ar, 10% $H_2$ and 0.1% $CH_4$. The CVD chamber is cooled to below 120 °C before venting in order to prevent excessive oxidation of Cu.

After annealing, the polished Cu foils (Alfa-aesar, 99.8% purity) present few large Cu grains (up to ~1 $cm^2$). These domains have preferential Cu(111) crystal orientation, but no rotational alignment, as suggested by LEED measurements in our previous work [18], as well as by other groups [S1]. By comparing different grains, we observe rotational misalignment of the large single-layer crystals, which, however, maintain the exclusive 0° or 30° relative orientation to the second layer.

**Semi-dry transfer from Cu to $SiO_2$/Si**

We spin coat the Cu foil with PMMA and cure it at 90 °C for 2 minutes. We then place a rigid PDMS frame on top of it and proceed to electrochemical delamination in NaOH 1M aqueous solution. Pt is used as counter-electrode, with a voltage of 2.4 V applied. A typical current of 0.8 mA results in delamination of the graphene/PMMA membrane in approximately 10 minutes. The stack is handled with tweezers via the supporting PDMS frame and rinsed in DI water for 15 minutes. A home-built transfer setup is used to deterministically place the graphene/PMMA stack in contact with a $SiO_2$/Si wafer, heated at 90 °C. We heat the sample further to 105 °C for 5 minutes, to promote the adhesion of the graphene crystals. The PDMS frame is then peeled off and the PMMA removed in acetone over 3 hours.

**Encapsulation in hBN**

We exfoliate thin hBN flakes from bulk single crystals on $SiO_2$/Si. Flat and uniform flakes are selected on the basis of their optical contrast, dark-field imaging and AFM. We first pick-up a hBN flake using a stamp consisting of a PC film on top of a PDMS block, supported by a glass slide, with the substrate heated to 60 °C. The flake is then placed in contact with a QC-TBG on $SiO_2$/Si for 5 minutes at 60 °C and peeled off, resulting in pick-up of the hBN-covered part. The graphene/hBN stack is then put in contact to the bottom hBN flake, which is picked-up as before, completing the encapsulation. Finally, we released the hBN/QC-TBG/QC stack at 160 °C on a $SiO_2$/Si with pre-patterned alignment markers, by slowly moving the contact front over the heterostructure to promote interface cleaning. The sample is finally rinsed in chloroform for 15 minutes.

**TEM**

Selected Area Electron Diffraction (SAED) was performed on a JEOL 2010F TEM at 80kV accelerating voltage. SAED patterns were collected from approximately 2 μm diameter regions of suspended graphene with camera length 50 cm.

**Raman spectroscopy**

We perform Raman spectroscopy with a commercial confocal system, using a 532 nm wavelength laser and 1 mW power. The spectra are calibrated using the Si peak at 520 cm$^{-1}$. The scanning Raman maps are obtained using a 100x objective, giving a 1 μm spot size, corresponding to the step size employed.

**Device fabrication**

The hBN/QC-TBG/hBN stacks are processed using standard e-beam lithography, combined with reactive ion etching and metal evaporation. We first fabricate a top-gate (Cr/Au 5/50 nm), ensuring to avoid contact to exposed graphene edges. We then pattern the electrical contacts, etch the heterostructure in $CF_4/O_2$ and evaporate (Cr/Au 5/50 nm) edge contacts in a single lithographic step; this minimizes contamination of the graphene edges and improve the device performance. Finally, we etch the Hall bar shape, using the physical mask provided by the metallic top gate, in combination with additional e-beam-patterned PMMA arms. The devices are then mounted on dual-in-line chip carriers and wire-bonded using Al wires.

**Low-temperature magnetotransport**

We study the devices in a $^4$He cryostat with a 12 T superconducting coil. We use standard AC lock-in detection (13 Hz) with a constant 10 nA current through the device. The top and bottom gates are biased via two independent channels of a source-meter and controlled via a home-built acquisition program. The map in Figure 3b is acquired as repeated top-gate sweeps at fixed bottom-gate voltages. The maps in Figure 3e are obtained by combining the top and bottom gates as to keep a constant displacement field, while sweeping the carrier density at fixed magnetic fields.

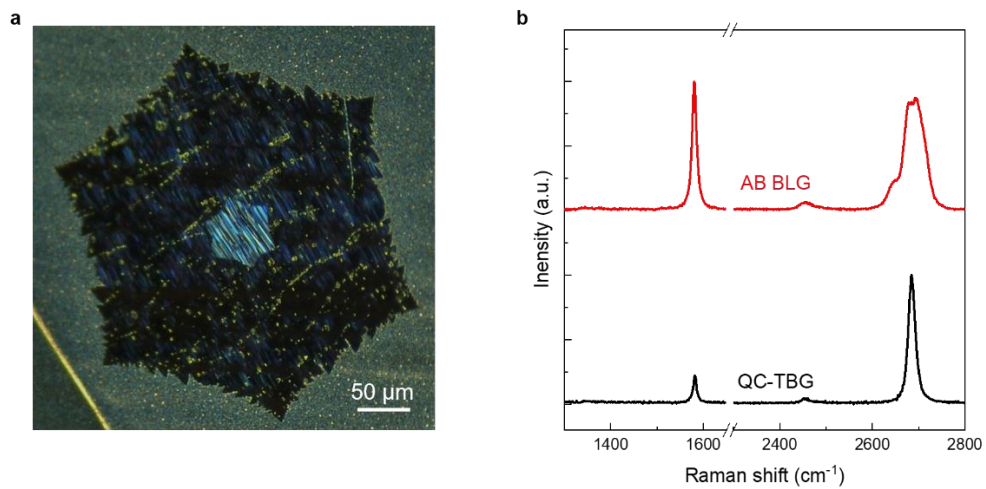

**Figure S1: Optical image and Raman of CVD AB bilayer graphene.** (a) Dark field optical microscopy image of an AB-stacked bilayer graphene crystal on Cu foil. (b) Representative Raman spectrum of AB bilayer graphene (red line), after transferring to SiO$_2$/Si; a spectrum of QC-TBG (black) is shown for comparison.

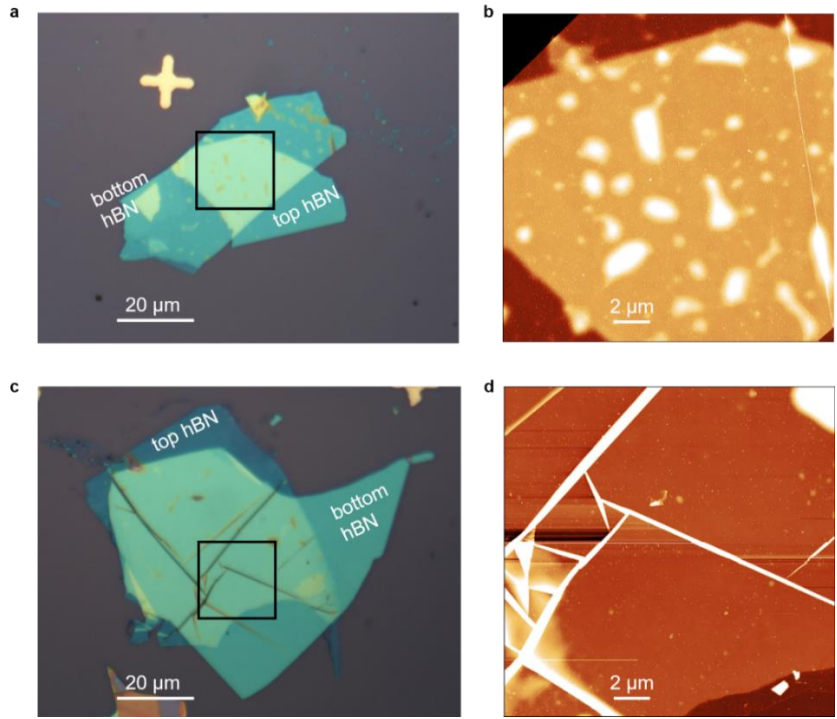

**Figure S2: Optical images and AFM of hBN-encapsulated samples.** (a) Optical image (100x magnification) of a hBN/QC-TBG/hBN with considerable interface contamination. (b) Atomic force microscopy (AFM) over the square area indicated in (a), highlighting numerous blisters. (c) Optical image (100x magnification) of a hBN/QC-TBG/hBN with successful interface cleaning. (d) AFM over the square area in (c), showing large and flat bubble-free areas.

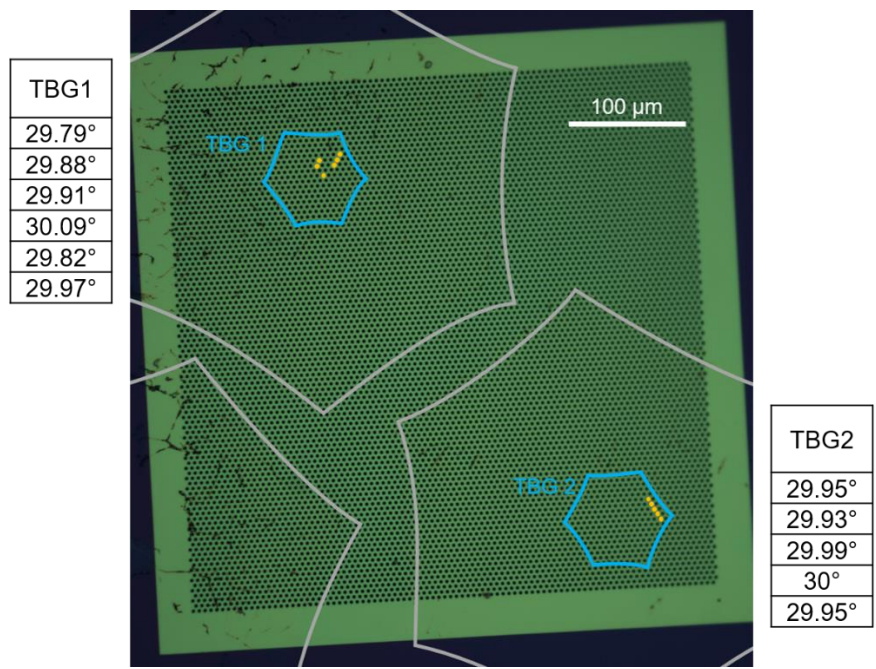

**Figure S3: Spatial homogeneity and experimental uncertainty of the twist angle.** Optical image of a TEM grid with suspended CVD graphene crystals (outlined by the grey lines), two of them showing large QC-TBG regions (blue lines). The yellow dots indicate the holes over which we measured SAED and determined the values of twist angle listed in the Figure. The twist angle was measured from SAED images by subtracting the average angular orientation of one layer's Bragg peaks from those of the other layer. An upper bound on uncertainty of 0.3° is estimated from a single layer Bragg peaks' standard deviation from 60°. This deviation could result from TEM imaging conditions such as slight tilt and astigmatism. However, it is likely the actual uncertainty is reduced when averaging multiple Bragg peaks in each image. Finally, combining the twist angle values obtained over the individual holes, we obtained a spatially-averaged twisting of 29.91° ± 0.11° for TBG1 and 29.96° ± 0.03° for TBG2.

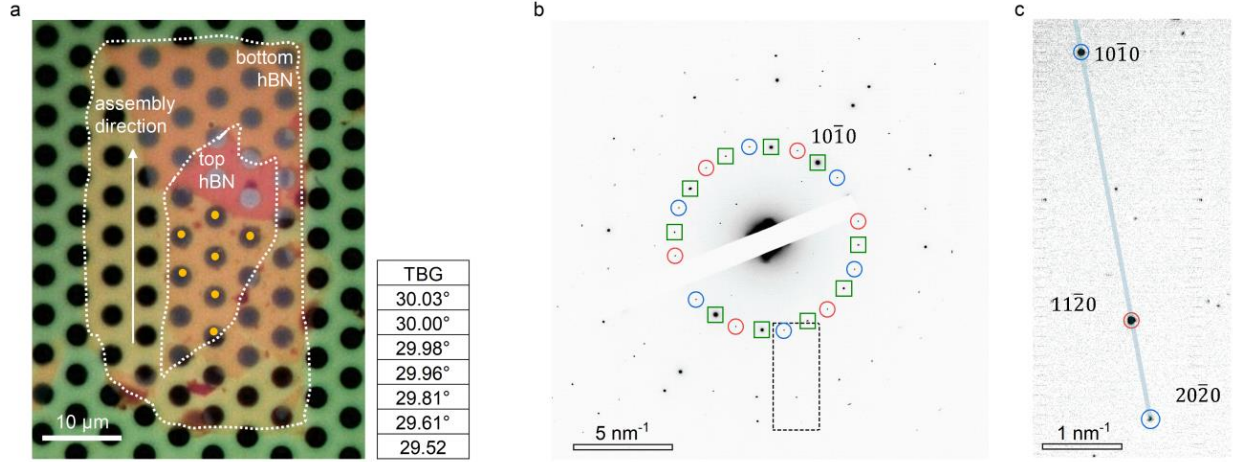

**Figure S4: SAED on hBN-encapsulated QC-TBG.** (a) Optical microscopy image of an hBN/QC-TBG/hBN sample on $Si_3N_4$ TEM grid. The dashed white lines outline the two hBN flakes. The ~50 nm thickness of the encapsulated part, comparable to that of devices D1,2, makes this sample representative of the devices used for the electrical transport experiments. The white arrow indicates the direction of the van der Waals assembly, which we performed parallel to well-defined straight edges of the hBN flakes, and intentionally misaligned with respect to those of the graphene layers. This strategy should minimize interlayer rotation due to the tendency of graphene to align to the hBN lattice [23]. The assembly technique (similarly to the release step in the semi-dry transfer used for the freestanding samples of Figure S3) relies on a wide and uniform contact front (perpendicular to the white arrow and proceeding along it) [22], which ensures a highly directional stress and minimizes induced torque and the possibility of interlayer rotation. (b) Representative SAED image of hBN-encapsulated QC-TBG. The image contains four sets of Bragg peaks, two from the twisted graphene bilayer (red and blue circles), two from the hBN flakes (green squares); the labels are placed on the $10\bar{1}0$ peaks. The rotational misalignment between the graphene layers and the hBN is evident. The hBN flakes are also misaligned to each other by approximately 30° due to different nature (zig-zag or armchair) of the reference straight edges. (c)

Zoomed in view of dashed rectangle in panel (b), highlighting the radial alignment between the $11\bar{2}0$ Bragg peaks from one graphene layer (red circle) and the $10\bar{1}0$ and $20\bar{2}0$ peaks from the other layer (blue circles). As in the case of freestanding QC-TBG (Figure 1e), this observation indicates 12-fold rotational symmetry. To improve the visibility of the $20\bar{2}0$ peak, the contrast of the image was increased with respect to panel (b). We measured analogous SAED patterns over the holes marked by the yellow dots in (a) and estimate the angle values listed in the table, with the same method used for the freestanding samples (see Figure S3, an upper bound of 0.3° applies on the measurement uncertainty). By combining the individual values, we obtain an average twist angle 29.84° ± 0.20°. The slightly larger deviation from 30 degrees in the encapsulated sample can be due to strain localized at interface bubbles, appreciable in (a). However, we did not find a correlation between the presence of bubbles in real space imaging and twist variations in SAED. Wrinkles in the heterostructure can form during the assembly and drive the angular distortion. In an additional sample made with thin hBN flakes (total thickness ~10 nm, not shown here), we observed a large ~1.5° change in the interlayer rotation across a macroscopic wrinkle. This might indicate that thin heterostructures (not representative of the devices used in the transport experiments), more prone to form wrinkles during the assembly, can be subjected to considerable twist angle distortions. In this sense, we stress that device D1 was fabricated on an accurately selected bubble- and wrinkle-free area (see Figure S5a) and it was surely not affected by the mechanisms discussed here. We also note that in Ref.[40] QC-TBG was placed on top of hBN, and, despite forming bubbles and wrinkles, a preserved 30° twisting was confirmed by atomically-resolved scanning probe measurements on flat areas.

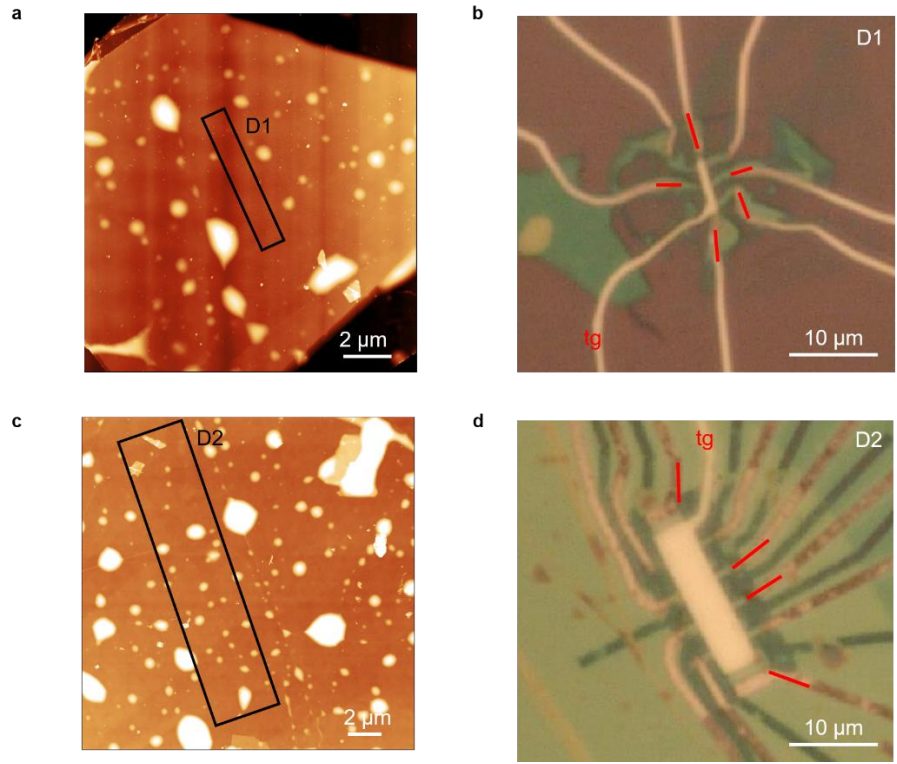

**Figure S5: Double gated Hall bar devices from CVD QC-TBG.** (a) AFM of the hBN/QC-TBG/hBN stack used to fabricate device D1. The 1×6 μm² rectangle indicates the blister-free area corresponding to the active channel of D1. (b) Optical microscopy image of device D1. The red lines mark the electrical contacts used for the measurements reported in the manuscript. (c) Same as (a), for device D2. The device area is 4×18 μm², comprising several residual blisters. (d) Same as (b), for device D2. The contacts to the top gate (tg) are designed to cross the edge of the heterostructure over graphene-free areas (individuated by Raman mapping), avoiding electrical shortage to the device channels.

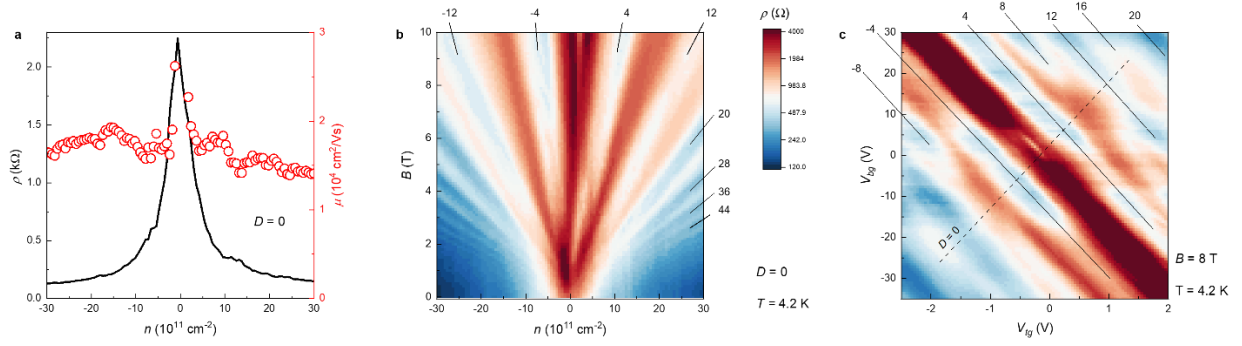

**Figure S6: Low-temperature transport data from device D2.** (a) Resistivity of device D2 as a function of carrier density at $D = 0$. The peak width is $n^* = 1.5 \times 10^{11}$ cm$^{-2}$ and the carrier mobility is 15-20 $\times 10^3$ cm$^2$/Vs. (b) Resistivity as a function of $n$ and $B$, at $D = 0$, showing a 8-fold degenerate Landau fan (in analogy to Figure 3e). (c) Resistivity as a function of the gate voltages at $B = 8$ T (same color scale as (b)). The filling factor sequence is 4, 12, 20, … close to $D = 0$, while additional states at 8, 16, 24, … appear at large displacement field. All data are acquired at 4.2 K.

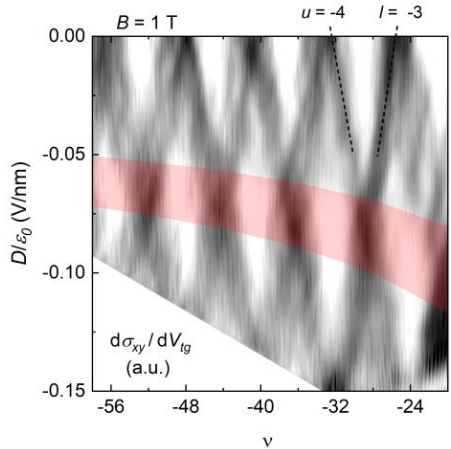

**Figure S7: LLs crossings in D1 and interlayer capacitance.** Zoom in main text Figure 3c, highlighting LLs crossings measured in D1 at $B = 1$ T. The displacement field $D$ results in a density mismatch $\Delta n$ and an electrostatic potential difference $\Delta V$ between the layers, according to $D =$

$e\Delta n/2 - C_{gg}\Delta V$, where $C_{gg}$ is the interlayer capacitance per unit area [6]. At the crossings, these quantities are defined by the LL index for the upper and lower layer (e.g. $u = -4$ and $l = -3$ for the one at $v = -28$), according to $\Delta n = (u-l)4eB/h$ and $-e\Delta V = (E_u-E_l)$, where $E_{u,l}$ are the corresponding LL energies. Therefore, $C_{gg}$ can be estimated from the positioning of each crossing. As the crossings extend over finite intervals due to level broadening, correspondingly, we estimate a finite range for the interlayer capacitance $C_{gg} = 6.9 \pm 1.4$ µF/cm² (red shadowed area), matching previous experimental findings on large-angle TBG [6,35].

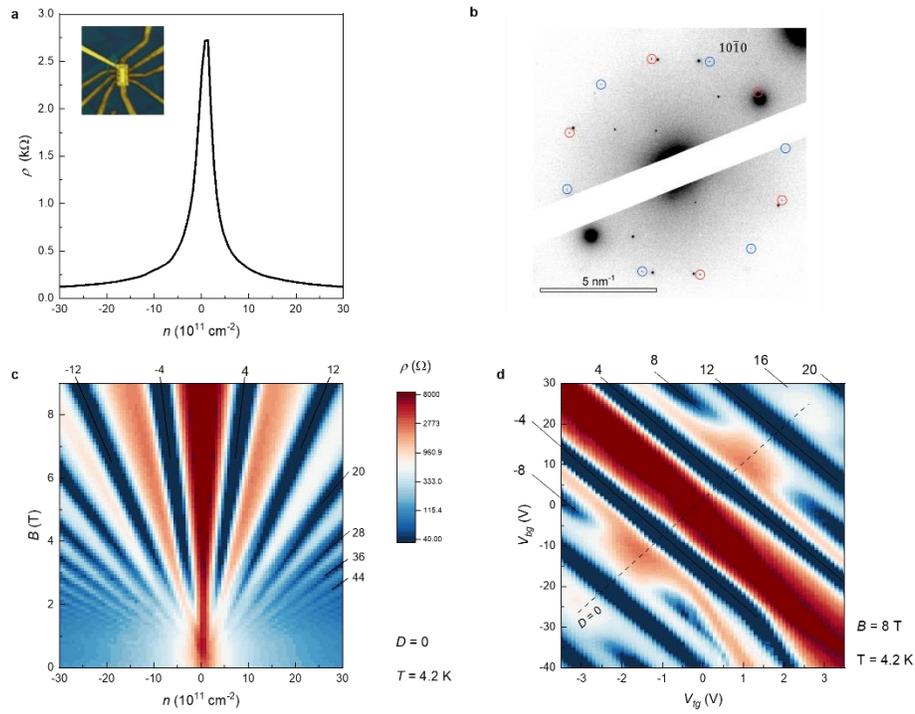

**Figure S8: Magnetotransport on device D3, TBG from exfoliated graphene flakes.** (a) Resistivity of device D3 as a function of carrier density at $D = 0$. The carrier mobility is ~40 x 10³ cm²/Vs. Inset; optical microscopy image of D3, made from hBN-encapsulated TBG obtained by artificial stacking of two exfoliated single-layer graphene. (b) Representative SAED image for device D3, acquired before device processing. The $10\bar{1}0$ peaks for the two graphene layers are marked by the red and blue circles. The twist angle, measured over 6 positions with the same method and experimental uncertainty as in Figure S3-S4, has an average value $30.22° \pm 0.09°$. (c) Resistivity as a function of n and B, at $D = 0$, showing an 8-fold degenerate Landau fan analogous to S6 (b). (d) Resistivity as a function of the gate voltages at $B = 8$ T. The x and y axis scales are matched to the ones of S6 (c) by taking into account the slight differences in gate capacitance and residual doping.

ADDITIONAL REFERENCES